\documentclass[twocolumn,aps,prl,showpacs,superscriptaddress]{revtex4-1}
\usepackage{graphicx}
\usepackage{amsmath,amssymb}
\usepackage{color}
\usepackage{multirow}

\newcommand{\ket}[1]{|#1\rangle}

\begin{document}


\title{Hybrid ququart-encoded quantum cryptography protected by Kochen-Specker contextuality}


\author{Ad\'an Cabello}
 \email{adan@us.es}
 \affiliation{Departamento de F\'{\i}sica Aplicada II, Universidad de
 Sevilla, E-41012 Sevilla, Spain}
 \affiliation{Department of Physics, Stockholm University, S-10691 Stockholm, Sweden}
 \author{Vincenzo D'Ambrosio}
 \affiliation{Dipartimento di Fisica della ``Sapienza''
 Universit\`{a} di Roma, I-00185 Roma, Italy}
 \author{Eleonora Nagali}
\affiliation{Dipartimento di Fisica della ``Sapienza''
 Universit\`{a} di Roma, I-00185 Roma, Italy}
\author{Fabio Sciarrino}
 \email{fabio.sciarrino@uniroma1.it}
 \homepage{http://quantumoptics.phys.uniroma1.it}
 \affiliation{Dipartimento di Fisica della ``Sapienza''
 Universit\`{a} di Roma, I-00185 Roma, Italy}
 \affiliation{Istituto Nazionale di
 Ottica, Consiglio Nazionale delle Ricerche (INO-CNR), I-50125 Florence, Italy}


\date{\today}



\begin{abstract}
Quantum cryptographic protocols based on complementarity are
nonsecure against attacks in which complementarity is imitated
with classical resources. The Kochen-Specker (KS) theorem
provides protection against these attacks, without requiring
entanglement or spatially separated composite systems. We
analyze the maximum tolerated noise to guarantee the security
of a KS-protected cryptographic scheme against these attacks,
and describe a photonic realization of this scheme using hybrid
ququarts defined by the polarization and orbital angular
momentum of single photons.
\end{abstract}


\pacs{03.67.Dd, 03.65.Ud, 42.50.Xa, 42.50.Tx}


\maketitle


{\em Introduction. }Quantum key distribution (QKD) protocols
allow two distant parties to share a secret key by exploiting
the fundamental laws of quantum mechanics. However, standard
quantum cryptographic protocols based on quantum
complementarity, such as the Bennett-Brassard 1984 (BB84)
protocol \cite{BB84}, are not secure against attacks in which
the adversary imitates complementarity with classical resources
\cite{Svozil06}. Interestingly, BB84-like protocols can be
improved to assure ``the best possible protection quantum
theory can afford'' \cite{Svozil06} by exploiting the fact that
the Bell \cite{Bell64} and Kochen-Specker (KS) \cite{KS67}
theorems show that the outcomes of quantum measurements do not
admit local and noncontextual descriptions, respectively. The
extra security provided by the Bell theorem has been
extensively investigated \cite{Ekert91, BHK05, ABGMPS07}.
However, this extra security is based on the assumption that
the legitimate parties can perform a loophole-free Bell test,
something which is beyond the present technological
capabilities and is not expected to be an easy task in the
future \cite{R09}. A similar problem affects recent proposals
combining the KS theorem with entanglement \cite{Cabello10,
HHHHPB10}. Therefore, it is worth exploring the extra security
offered by the KS theorem in situations which require neither
entanglement nor composite systems, but only single systems
with three or more distinguishable states. For cryptographic
purposes, the difference between qubits and systems of higher
dimensionality is this: Whereas in qubits different bases are
always disjoint, from qutrits onward different bases may share
common elements. It is this property which is at the root of
the proofs of Bell and KS theorems.

Here we investigate the experimental requirements for obtaining
the extra security offered by a KS-protected QKD protocol
introduced by Svozil \cite{Svozil09}, based on the properties
of the simplest KS set of states \cite{CEG96}. Hence we propose
to implement such a protocol by adopting ququart states encoded
in the hybrid polarization-orbital angular momentum
four-dimensional space of single photon states \cite{Naga10pra,
Naga10prl}. For this purpose, we introduce the optical schemes
to measure all the states needed to prove KS contextuality. The
capability of encoding a four-dimensional quantum state in a
single photon by exploiting these two different degrees of
freedom enables us to achieve a high stability and transmission
rate in free-space propagation.


\begin{figure}[t!]
 \centering	
 \includegraphics[width=8.2cm]{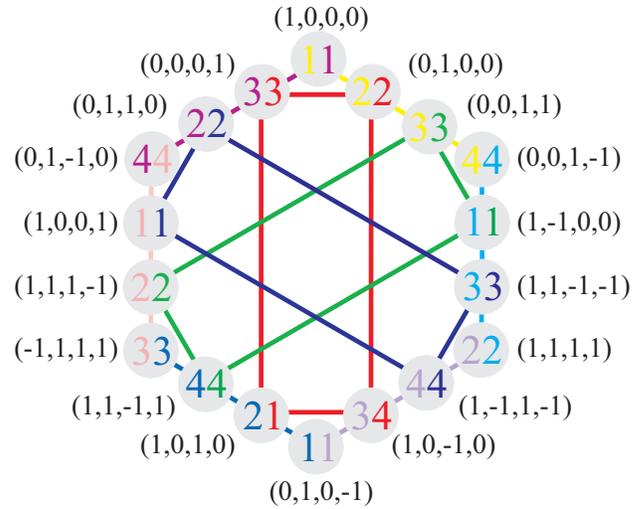}
 \caption{(Color online) The protocol is based on a KS set of 18 states which can be grouped in
 9 bases represented by 9 colors. Every state belongs to two different bases.
 No set of 18 balls can have all the properties required to imitate the KS set;
 at least two balls must have different symbols. Therefore, the imitation can be detected.}
 \label{KS18}
\end{figure}


{\em Svozil's protocol. }The cryptographic protocol introduced
by Svozil in \cite{Svozil09} is a variation of the BB84
protocol and works as follows: (i) Alice randomly picks a basis
from the nine available in Fig.~\ref{KS18} and sends Bob a
randomly chosen state of that basis. (ii) Bob, independently
from Alice, picks a basis at random from the nine available and
measures the system received from Alice. (iii) Bob announces
his bases over a public channel, and Alice announces those
events in which the state sent belongs to the measured basis.
Therefore, the probability of adopting the same basis is
$\frac{1}{9}$. (iv) Alice and Bob exchange some of the
remaining matching outcomes over a public channel to ensure
that nobody has spied their quantum channel. (v) Alice and Bob
encode the four outcomes by using four different symbols. As a
result, for each successful exchange Bob and Alice share a
common random key.

The advantage of this protocol over the BB84 protocol is that
it is protected by the KS theorem against attacks in which the
adversary replaces the quantum system with a classical one.
These attacks can be described using a classical toy model
\cite{Svozil06, Svozil09} in which, in step (i), Alice is
actually picking one of nine differently colored eyeglasses
(instead of one of the nine different bases in Fig.~\ref{KS18})
and picking a ball from an urn (instead of picking one of the
18 states in Fig.~\ref{KS18}) with two color symbols in it
(corresponding to the two bases the state belongs to). Each one
of the 9 differently colored eyeglasses allows her to see only
one of the nine different colors. To reproduce the quantum
predictions: $(a)$ each of the balls must have one symbol $S_i
\in \{1,2,3,4\}$ written in two different colors chosen among
the 18 possible pairs. Her choice of eyeglass decides which
symbols Alice can see. $(b)$ All colors are equally probable
and, for a given color, the four symbols are equally probable.
In step (ii), Bob is actually picking one of nine differently
colored eyeglasses and reading the corresponding symbol. A
classical strategy like this one can successfully imitate the
quantum part of the BB84 protocol (see \cite{Svozil06} for
details) but not the protocol described above. The reason is
that the requirements $(a)$ and $(b)$ cannot be satisfied
simultaneously. Figure~\ref{KS18} shows how to prepare 18 balls
with the minimum number of balls not having the same symbol.


{\em Experimental requirements. }As shown in Fig.~\ref{KS18},
the minimum number of balls not having the same symbol is two
out of 18. A ball attack can be detected only in those runs in
which Alice and Bob pick differently colored eyeglasses.
Therefore, for the set in Fig.~\ref{KS18}, the trace of such an
attack will be a $\frac{2}{18}$ probability of Alice picking a
symbol such that the corresponding interlinked symbol (seen
only with differently colored eyeglasses) is different. As a
consequence, to demonstrate that the experimental results
cannot actually be imitated with balls and to experimentally
certify the extra security of this KS-based QKD protocol, we
need an experimental probability $w$ of wrong state
identification, defined as the probability that Bob makes a
wrong identification of the state sent by Alice when Bob has
successfully measured in a correct basis, of $w < \frac{1}{9}
\approx 0.111$.


\begin{table*}[ht]
\begin{tabular}{c l c c c c c c c}
\hline\hline
Set & Logic & E & S & PSI1 & PSI2 & QWP & MC & WP \\
\hline\hline
I & $(1,0,0,0),(0,1,0,0)$  & & b & & & & $\frac{\pi}{4}$ & \\
  & $(0,0,1,1),(0,0,1,-1)$ & & b & & & & $\frac{\pi}{4}$ & \\
II & $(1,1,1,1),(1,1,-1,-1)$ & & c & & & & $\frac{\pi}{4}$ & H $\frac{\pi}{8}$ \\
   & $(1,-1,0,0),(0,0,1,-1)$ & & c & & & & $\frac{\pi}{4}$ & H $\frac{\pi}{8}$ \\
III & $(1,1,1,1),(1,-1,1,-1)$ & & b & & & & $\frac{\pi}{4}$ & \\
    & $(1,0,-1,0),(0,1,0,-1)$ & & b & & & & $\frac{\pi}{4}$ & \\
IV & $(-1,1,1,1),(1,1,-1,1)$ & $\checkmark$ & b & $\checkmark$ & & $\frac{\pi}{4}$ & 0 & \\
   & $(1,0,1,0),(0,1,0,-1)$  &              & b & $\checkmark$ & & $\frac{\pi}{4}$ & 0 & \\
V & $(1,0,0,1),(0,1,-1,0)$  & $\checkmark$ & c & $\checkmark$ & $\checkmark$ & & $\frac{\pi}{4}$ & H $\frac{\pi}{8}$ \\
  & $(1,1,1,-1),(-1,1,1,1)$ & $\checkmark$ & c & $\checkmark$ & $\checkmark$ & & $\frac{\pi}{4}$ & H $\frac{\pi}{8}$ \\
VI & $(1,0,0,1),(0,1,1,0)$     & $\checkmark$ & c & & $\checkmark$ & & 0 & Q $\frac{\pi}{4}$ \\
   & $(1,1,-1,-1),(1,-1,1,-1)$ & & c & & $\checkmark$ & & 0 & Q $\frac{\pi}{4}$ \\
VII & $(1,1,1,-1),(1,1,-1,1)$ & $\checkmark$ & c & & $\checkmark$ & & 0 & Q $\frac{\pi}{4}$ \\
    & $(0,0,1,1),(1,-1,0,0)$  & & c & & $\checkmark$ & & 0 & Q $\frac{\pi}{4}$ \\
VIII & $(0,0,0,1),(1,0,1,0)$  & & c & & & & & H $\frac{\pi}{8}$ \\
     & $(1,0,-1,0),(0,1,0,0)$ & & c & & & & & H $\frac{\pi}{8}$ \\
IX & $(0,1,-1,0),(0,1,1,0)$ & $\checkmark$ & b & $\checkmark$ & & $\frac{\pi}{4}$ & 0 & \\
   & $(1,0,0,0),(0,0,0,1)$  & & b & $\checkmark$ & & $\frac{\pi}{4}$ & 0 & \\
\hline\hline
\end{tabular}
\caption{\label{Table}The ququart states that compose the KS
set are divided in nine basis and encoded in polarization and
orbital angular momentum degrees of freedom by adopting the
devices in Fig.~\ref{setup}. Column E identifies whether or not
the states are entangled states of the two degrees of freedom.
Column S specifies the experimental setup to be adopted for the
analysis. Column MC gives the angle between the horizontal axis
and the orientation of the cylindrical lenses of the mode
converter MC. Column QWP indicates the angle of the quarter
wave plate to be inserted after the PSI in Fig.~\ref{setup}-b
for the analysis of bases IV and IX.WP refers to the type of
wave plate to be inserted in the setup: H means half-wave plate
and Q means quarter-wave plate.}
\end{table*}


\begin{figure*}[ht]
 \centering	
 \includegraphics[width=14.6cm]{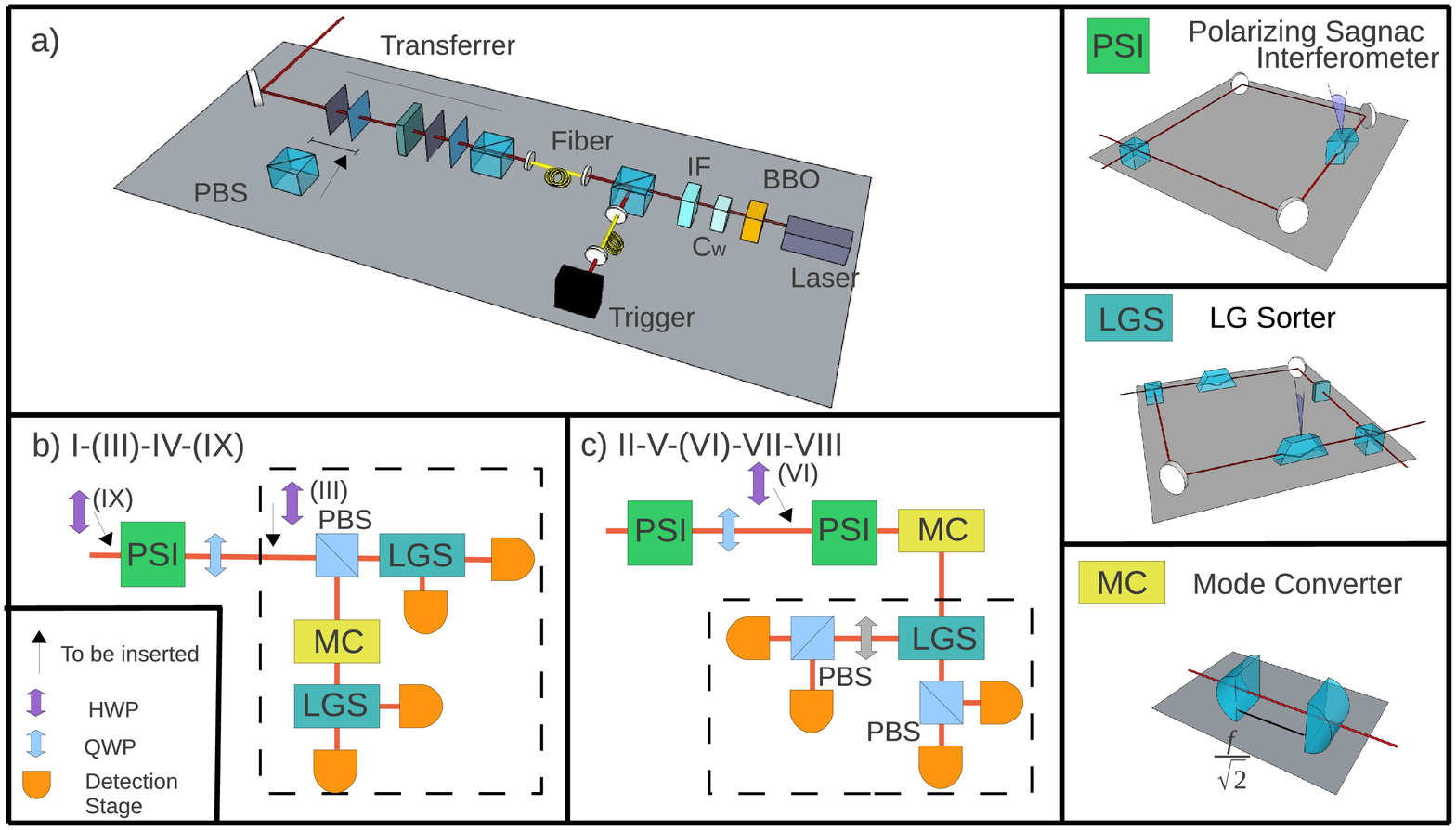}
 \mbox{}\vspace{-1cm}
\caption{(Color online) (a) Setup for the generation of ququart
states: One of the two photons emitted by SPDC acts as a
trigger, while the other one is sent to a polarizing beam
splitter (PBS), wave plates and a quantum transferrer based on
the $q$-plate in order to generate the desired ququart. (b)
Setup for the analysis of bases (I-III-IV-IX): The setup in the
dotted rectangle analyzes the four states of basis I; basis III
can be measured by inserting a half-wave plate (HWP) at $\pi/8$
before the PBS. A polarizing Sagnac interferometer (PSI) and a
quarter-wave plate are needed to analyze bases IV and IX
(adding a HWP at $\pi/8$ before the PSI). (c) Setup for the
analysis of bases (II-V-VI-VII-VIII): The part in the dotted
rectangle is suitable to sort the four states of all the bases
(the gray wave plate can be a HWP or a QWP depending on the
particular basis as shown in Table~\ref{Table}); this part is
sufficient to analyze basis VIII. Basis II can be analyzed by
adding a mode converter (MC). Using a PSI before the MC makes
it possible to analyze bases VI (adding a HWP at $\pi/8$) and
VII. Finally, the states of basis V can be sorted by an
additional PSI and QWP. The pictures in the three boxes on the
right represent the Sagnac interferometer, the LG mode sorter,
and the cylindrical lens mode converter, respectively. The
detection stage consists of a $q$-plate, a single-mode fiber,
and a detector.} \label{setup}
\end{figure*}


{\em Implementation using polarization- and orbital angular
momentum-encoded ququarts. }Here we propose a scheme for the
experimental implementation of the KS-protected QKD protocol.
To test its feasibility, we need to prepare the 18 states,
measure each of them in two different bases, and obtain an
average value of $w$ over the $18 \times 2$ possibilities. The
condition which must be fulfilled is $w < 0.111$, which
corresponds to a mean fidelity value of the transmission of the
state of $F=0.889$. In addition, to check that any intercept
and resend strategy causes a disturbance, one should be able to
measure what happens when the states are measured in the wrong
basis. While in the correct basis the probabilities for the
four possible outcomes are (in the ideal case) $0$, $0$, $0$,
and $1$, in the wrong basis they are either $0$, $0$,
$\frac{1}{2}$, and $\frac{1}{2}$ or $0$, $\frac{1}{4}$,
$\frac{1}{4}$, and $\frac{1}{2}$.

Svozil's protocol uses nine sets of four-dimensional states
defining a 18-state KS set. We propose encoding
four-dimensional quantum states by exploiting two different
degrees of freedom of the same particle, an approach that
allows us to achieve higher efficiency in the transmission
process. It has recently been demonstrated that ququart states
can be efficiently generated by manipulating the polarization
and orbital angular momentum (OAM) of a single photon
\cite{Naga10pra}. In particular, we consider a bidimensional
subset of the infinite-dimensional OAM space, denoted as $o_1$,
spanned by states with OAM eigenvalue $m=\pm 1$ in units of
$\hbar$. According to the nomenclature
$\ket{\varphi,\phi}=\ket{\varphi}_{\pi}\ket{\phi}_{o_1}$, where
$\ket{\cdot}_{\pi}$ and $\ket{\cdot}_{o_1}$ stand for the
photon quantum state ``kets'' in the polarization and OAM
degrees of freedom, the logic ququart basis can be rewritten
as:
\begin{equation}
\{\ket{1},\ket{2},\ket{3},\ket{4}\}\rightarrow\{\ket{H,+1},\ket{H,-1},\ket{V,+1},\ket{V,-1}\},
\end{equation}
where $H$ ($V$) refers to horizontal (vertical) polarization.
Following the same convention, the OAM equivalent of the basis
$\ket{H}$ and $\ket{V}$ is then defined as
$\ket{h}=\frac{1}{\sqrt{2}}(\ket{+1}+\ket{-1})$ and
$\ket{v}=\frac{i}{\sqrt{2}}(\ket{+1}-\ket{-1})$. Finally, the
$\pm45^{\circ}$ angle ``antidiagonal'' and ``diagonal'' linear
polarizations are hereafter denoted by the kets
$\ket{A}=(\ket{H}+\ket{V})/\sqrt{2}$ and
$\ket{D}=(\ket{H}-\ket{V})/\sqrt{2}$, while the OAM equivalent
is denoted by $\ket{a}=(\ket{h}+\ket{v})/\sqrt{2}$ and
$\ket{d}=(\ket{h}-\ket{v})/\sqrt{2}$. It is convenient to work
with Laguerre-Gauss laser modes (LG$_{0,\pm 1}$) as OAM
eigenstates since, in this case, the states
($\ket{h}$,$\ket{v}$,$\ket{a}$,$\ket{d}$) will result as the
Hermite-Gauss modes (HG$_{1,0}$, HG$_{0,1}$) along the axes and
rotated by $45^{\circ}$. This feature allows us to easily
transform the states by an astigmatic laser mode converter
\cite{Beij93opt}. We stress that by choosing a bidimensional
subspace of OAM we avoid detrimental effects on the state due
to the radial contribution in the free propagation and
Gouy-phases associated with different OAM values
\cite{Naga09opt}. Hence a hybrid approach for the encoding of a
ququart state, based on OAM and polarization, leads to a higher
stability for the single photon propagation compared to a qudit
implemented only by adopting the OAM degree of freedom.
According to the previous definitions, a state $(a_1, a_2, a_3,
a_4)$ of the KS set is implemented as
\begin{equation}
a_1\ket{H,+1}+a_2\ket{H,-1}+a_3\ket{V,+1}+a_4\ket{V,-1}.
\end{equation}
The coefficients $a_i$ for each state are shown in
Table~\ref{Table}, along with the settings needed to analyze
each basis.


{\em Generation. }Figure~\ref{setup} shows the optical schemes
for the generation and detection of any ququart state of the KS
set. The generation of the states can be achieved by adopting a
spontaneous parametric down conversion (SPDC) source of pair of
photons, as in Fig.~\ref{setup}(a), where we consider a
collinear generation of couples $\ket{H}\ket{V}$, where one of
the two photons acts as a trigger for the heralded generation
of a single photon to be sent to the experimental setup. As in
\cite{Naga10pra}, the manipulation of the OAM degree of freedom
can be achieved by adopting the $q$-plate device
\cite{Naga09prl, Naga09opt}. On the polarization, the $q$-plate
acts as a half-wave plate, while on the OAM it imposes a shift
on the eigenvalue $m=\pm2q$, where $q$ is an integer or
half-integer number determined by the (fixed) pattern of the
optical axis of the device. In order to manipulate the OAM
subspace $o_1=\{\ket{+1},\ket{-1}\}$, a $q$-plate with
topological charge $q=1/2$ should be adopted \cite{Sluss11}.
Interestingly, the fact that the $q$-plate can entangle or
disentangle the OAM and polarization degrees of freedom can be
exploited for the preparation of any ququart states. In order
to generate all the states of the KS set, it is sufficient to
exploit a technique based on a quantum transferrer
$\pi\rightarrow o_1$ described in \cite{Naga10pra}. The OAM
eigenmodes produced in this way are not exactly LG modes but
hypergeometric Gaussian ones \cite{Kar07}. Since some of the
detection schemes are based on the properties of
Laguerre-Gaussian modes, this fact will lead, in some cases, to
a detection efficiency of around 80\%. Thus, in order to avoid
noise due to different OAM order contributions, it is
sufficient to insert in the detection stage a $q$-plate and a
single-mode fiber connected to the detector (see
Fig.~\ref{setup}).


{\em Measurement of the KS bases. }The bases involved in the KS
set have different structures as shown in Table~\ref{Table}.
They can be classified in three groups, depending on whether
they are composed of separable, entangled (between polarization
and OAM) or both separable and entangled states.

The detection setup is shown in Figs.~\ref{setup}(b) and (c).
Their components are a polarizing Sagnac interferometer with a
Dove prism (PSI) \cite{SDPMS10}, an astigmatic laser mode
converter (MC) \cite{Beij93opt}, and a Laguerre-Gauss mode
sorter (LGS) \cite{Wei03}. The PSI consists of a Sagnac
interferometer with a polarizing beam splitter as input-output
gate and a Dove prism that intercepts the two
counterpropagating beams and can be rotated around the optical
axes. This scheme allows us, under appropriate conditions, to
transform an entangled state into a separable one. In this
case, the prism must be rotated in order to add a phase shift
of $\Delta\phi=\pi/2$ between $\ket{H}$ and $\ket{V}$
($\alpha=\pi/8$ in Fig.~\ref{setup}). For example, the states
of basis IV are transformed into
($\ket{L,a},\ket{L,d},\ket{R,+1},\ket{R,-1}$). The MC consists
of two cylindrical lenses (with the same focal length $f$) at
distance $f/\sqrt{2}$. It allows us to convert the HG states
($\ket{a}$,$\ket{d}$) into ($\ket{+1}$,$\ket{-1}$) and, if
rotated by 45$^\circ$ along the optical axes, to convert
($\ket{h}$, $\ket{v}$) into ($\ket{+1}$, $\ket{-1}$)
\cite{Beij93opt}. The LGS consists of a Mach-Zehnder
interferometer with a Dove prism in each arm. The two prisms
are rotated by $\beta=\pi/4$ with respect to each other. A
phase plate ($\psi=\pi/2$) in one of the two arms allows us to
send $\ket{+1}$ and $\ket{-1}$ in the two different output
ports of the Mach-Zehnder. States belonging to sets
$I-III-IV-IX$ can be analyzed by adopting the scheme reported
in Fig.~\ref{setup}(b) with some slight modifications related
to the specific basis to be measured. The scheme in
Fig.~\ref{setup}(c) leads to the analysis of bases
$II-V-VI-VII-VIII$. All the details on the settings of the
different measurement devices are in Table~\ref{Table}.


{\em Conclusions. }Device-independent QKD based on
loophole-free Bell tests are still far in the future. It is
therefore worth investigating whether quantum contextuality can
produce some extra protection to BB84-like protocols which do
not use entangled states. Here we have presented a proposal to
demonstrate a quantum contextuality-based extra protection
against a particular attack, requiring neither composite
systems nor entangled states.


\begin{acknowledgments}
We acknowledge helpful discussions with E. Karimi and L.
Marrucci. This work was supported by the MICINN Project
No.~FIS2008-05596, the Wenner-Gren Foundation, FIRB Futuro in
Ricerca-HYTEQ, and Project PHORBITECH of the Future and
Emerging Technologies (FET) program within the Seventh
Framework Programme for Research of the European Commission,
under FET-Open Grant No.~255914.
\end{acknowledgments}



\end{document}